\documentclass[preprint,12pt]{article}
\usepackage{amssymb}
\usepackage{graphicx}
\usepackage{multirow}
\usepackage{caption}
\usepackage{subfig}
\begin{document}

\title
{Revisiting some physics issues related to the new mass limit for magnetized
white dwarfs}

\author{Upasana Das and Banibrata Mukhopadhyay \\
Department of Physics, Indian Institute of
  Science, Bangalore 560012, India\\
upasana@physics.iisc.ernet.in , bm@physics.iisc.ernet.in}

\maketitle
\begin{abstract}
We clarify important physics issues related to the recently established new mass 
limit for magnetized white dwarfs which is significantly super-Chandrasekhar. 
The issues include, justification of high magnetic field and the corresponding 
formation of stable white dwarfs, contribution of the magnetic field 
to the total density and pressure, 
flux freezing, variation of magnetic field and related currents therein.
We also attempt to address the observational connection of such highly 
magnetized white dwarfs.

\vskip0.5cm
{\it PACS Number(s)}: 97.20.Rp, 97.60.Bw, 97.10.Ld, 71.70.Di, 51.30.+i
\vskip1cm

\end{abstract}

\section{Introduction}

Recently, Mukhopadhyay and his collaborators have proposed that highly magnetized white dwarfs 
could have significantly super-Chandrasekhar masses \cite{kundu,prd12,ijmpd,prl13,apjl13,grf13}. 
In order to arrive at this result, the authors have made certain assumptions 
for the convenience of calculation. These are, the choice of constant magnetic field in 
the central region of white dwarfs, which decreases with the decreasing density near the surface of the 
white dwarf;
the choice of spherical symmetry for strongly 
magnetized white dwarfs; separate central and surface magnetic flux conservation
as the white dwarfs evolve (e.g., due to accretion); etc.

Now, in the present paper, we plan to justify the above assumptions in a greater detail, in 
the light of various related physics issues. This is particularly important, as 
several friends, colleagues and critics have raised certain questions about 
our proposal of significantly super-Chandrasekhar white dwarfs and a new mass limit of white dwarfs. 
This paper is by and 
large a suitable compilation of our responses to more than a dozen referee reports, which we 
have successfully tackled during this series of works starting with the one that
appeared in 2012 \cite{kundu}.

\section{Equipartition magnetic field for super-Chandrasekhar white dwarfs}
\label{equi}

In this section, we would like to clarify that for the super-Chandrasekhar white 
dwarfs considered in our works, the equipartition magnetic field ($B$) is 
no longer same as that for weakly magnetized white dwarfs, but much higher. 

One can estimate the equipartition field ($B_{\rm equi}$) of a star by recalling the scalar 
virial theorem. 
The gravitational potential energy of a spherical star which 
can be described by a polytropic equation of state (EoS), i.e., $P=K\rho^{\Gamma} = K\rho^{1+1/n}$ 
(where $P$ is the pressure, $\rho$ the mass density, $K$ a dimensional constant and $\Gamma$ the polytropic index), 
is given by $-\frac{3}{5-n}(GM^2/R)$ \cite{shapiro}.
By equating the magnetic and the gravitational 
potential energies of a spherical star of mass $M$ and radius $R$, i.e., 
$(4\pi R^{3}/3)(B^2/8\pi)= \frac{3}{5-n}(GM^{2}/R)$, one arrives at the 
maximum possible $B$, i.e., $B_{\rm equi}$, for the star. 
For a typical white dwarf having radius $10^4$ km and mass 
$1.4 M_{\odot}$, characterized by a polytropic EoS with $\Gamma=4/3$ or $n=3$, 
$B_{\rm equi}$ turns out to be $2.2\times 10^{12}$ G. 
However, as shown in 
\cite{prl13}, in the presence of a strong magnetic field and high Fermi energy ($E_F \gg m_ec^2$, 
when $m_e$ is the rest mass of electrons and $c$ the speed of light), 
the electron degenerate gas is Landau quantized and is characterized by a polytropic EoS with $\Gamma=2$ or $n=1$. 
Thus, the equipartition magnetic field for such a white dwarf is given by
$B_{\rm equi} =  2.265\times 10^{8} (M/M_{\odot})(R/R_{\odot})^{-2}$ G, where 
$M_\odot$ and $R_\odot$ are the mass and radius of Sun respectively and $G$ is 
Newton's gravitation constant.
Hence, for a super-Chandrasekhar white dwarf having, for example, mass $2.58 M_{\odot}$ and 
radius $70$ km, corresponding to a central magnetic field $ B_{\rm cent}=8.8\times 10^{17}$ G
(see, e.g., Figure 1 of \cite{apjl13}), $B_{\rm equi}~ 
= 5.77\times 10^{16}$ G $\gg 10^{12}$ G. 
Now this $B_{\rm equi}$ has to be compared with the average field ($B_{\rm avg}$), as obtained by us, 
for the white dwarf under consideration.
In order to estimate $B_{\rm avg}$, we define 
$B_{\rm avg} \approx (B_{\rm cent} + B_{\rm surf})/2$, where $B_{\rm surf}$ denotes the 
surface magnetic field. The combined gas and magnetic pressure needs to vanish at the surface of the 
white dwarf as per the theory of stellar structure. 
This is maintained as long as $B_{\rm surf}$ is several orders of magnitude 
lower than $B_{\rm cent}$, such that the surface magnetic pressure 
is much less compared to the central magnetic pressure. 
If this is true, which indeed we have always argued for 
in our works \cite{kundu,prd12,ijmpd,prl13,apjl13,grf13}, then one can easily see that 
$B_{\rm avg} \approx B_{\rm cent}/2$. 
Thus, for the super-Chandrasekhar white dwarf having mass $2.58M_\odot$ and 
$B_{\rm cent}=8.8\times 10^{17}$ G, we obtain 
$B_{\rm avg}=4.4\times10^{17}$ G.
Note that, although a large $B$ brings in anisotropy
in the system, which may lead the star to become a spheroid, making a general relativistic treatment
indispensable, in our works \cite{kundu,prd12,ijmpd,prl13,apjl13,grf13} we stuck to the Newtonian framework assuming the star
to be spherical. Therefore, in that case, the overall isotropic magnetic pressure would be $P_B=B^2/24\pi$.
This increases $B_{\rm equi}$ to $10^{17}$ G.
Therefore, apparently, the above mentioned $2.58M_\odot$ white dwarf corresponds to 
a $B_{\rm avg}$ which is same order of magnitude as $B_{\rm equi}$, but just $4.4$ times larger.
We will address the issue of dynamical stability of such a white dwarf in detail in \S\ref{stability}.

\section{Anisotropic effect in the presence of strong magnetic field}

If $B$ is very strong, then the pressure may become anisotropic \cite{sinha}
and, as shown by Bocquet et al. \cite{boquet}, such a high $B$ can lead to a flattening 
of the star due to the magnetic tension developed in the combined fluid-magnetic medium.
Such an effect could also occur in highly magnetized white dwarfs. 
In order to solve for the equilibrium structure of such a non-spherical white dwarf 
in the Newtonian framework itself, one needs to use methods such as that proposed by 
Ostriker and Hartwick \cite{ostriker}, known as the ``self-consistent-field" method.
For a general relativistic treatment, as stated by Paulucci et al. \cite{pauluci}, 
one has to generalize the Tolman-Oppenheimer-Volkoff
equation to take into account the anisotropy in the pressure, but to the best of 
our knowledge this has not been achieved so far, as is also mentioned 
in the same paper itself. 
We propose to carry out such calculations for highly magnetized white dwarfs in a future work. 
However, in the previous work \cite{prd12}, we have initiated the understanding 
of the effect of deviation from spherical symmetry due to the magnetic field, if any, in the Appendix. 
Interestingly we observe that flattening effects due to magnetic field renders 
super-Chandrasekhar white dwarfs even at relatively weaker magnetic fields which are 
more probable in nature.

\section{Justification of constant central magnetic field}

In our works, we have practically modeled the central region of the white dwarf 
having a strong constant magnetic field (several orders of magnitude larger 
than $B_{\rm surf}$). This can be justified as follows.

First, let us briefly recall how $B_{\rm cent}$ for our white dwarfs is obtained.
The maximum number of Landau levels ($\nu_m$) that can be occupied by a gas of electrons 
in a magnetic field, having maximum Fermi energy $E_{Fmax}$, is given by (see \cite{prd12} and also 
supplementary material of \cite{prl13})
\begin{equation}
\nu_m= \frac{\left(\frac{E_{Fmax}}{m_ec^2}\right)^{2} - 1}{2B_{D}} , 
\label{numax}
\end{equation}
where $B_D$ is $B$ in units of $B_c=4.414\times 10^{13}$ G. We note that
it is when $B_D \gg 1$, so that for a fixed $E_{Fmax}$ only a few Landau levels are occupied, 
the magnetic field plays an
important role in significantly modifying the EoS for the
relativistic electron degenerate gas (see Figure 1 of \cite{prd12}). 
We are interested in this very regime as it is in this case that the 
mass-radius relation of the underlying magnetized white dwarf gets appreciably modified to yield 
super-Chandrasekhar mass. In our works, we argue 
such a value of $B$ to be the $B_{\rm cent}$, 
as justified in \S IV.C of \cite{prd12}.
Note additionally that a high $E_F$ corresponds to a
large $\rho$, which in turn renders a large $B$ (from flux freezing theorem)
\cite{prl13}.
This assures $\nu_m=1$ at the center when both $E_{Fmax}$ and $B$ 
have their largest values, as seen from equation (\ref{numax}). 
The values of $B$ (or $B_{\rm cent}$)
for various combinations of $E_{Fmax}$ and $\nu_m$, as determined by equation (\ref{numax}), 
are given in Table I of \cite{prd12}.

Now, from Maxwell's equations in a steady state, we have
\begin{equation}
\nabla \times {\bf B} = \frac{4\pi}{c}{\bf j},
\label{current}
\end{equation}
where ${\bf j}$ is the current density. In the presence of a very strong central magnetic field, 
$B_{\rm cent} \gg B_c$ (as is the case for a 
significantly super-Chandrasekhar white dwarf), the electron degenerate 
matter will not only be Landau quantized but also the electrons will occupy mostly  
the ground Landau level, as explained above. Thus the electrons are not expected to move and hence practically there 
will be no current, i.e., ${\bf j}=0$. Therefore, from equation (\ref{current}), we obtain 
that there is no spatial variation of $B$, i.e., the magnetic field is constant. Since 
it is this $B_{\rm cent}$ which is primarily responsible for the mass to
exceed the Chandrasekhar limit (see, e.g., \S IV.C of \cite{prd12}), one can
consider the hydrostatic equilibrium condition in order to solve for the structure of the magnetized white 
dwarfs, at least in the central region, for obtaining super-Chandrasekhar white dwarfs.

We would like to point out here that, indeed, Ostriker and 
Hartwick \cite{ostriker} constructed models of white dwarfs 
with $B \gtrsim 10^{12}$ G at the center but with a much smaller 
field at the surface. Thus, the concept of high interior magnetic fields 
in white dwarfs, although difficult to verify observationally, are not implausible.

\section{Contribution of the magnetic field to the total density}
\label{rhoB}

Another commonly raised issue is the contribution of magnetic density ($\rho_B=B^2/8\pi c^2$) 
to the mass of the white dwarf.

Recall that $B$ inside the white dwarf appears to be constant in the 
central region and subsequently decreases with radius, 
which was already mentioned in previous papers \cite{prd12,prl13,apjl13}. We now point out that for 
$B_{\rm cent}<10^{17}$ G, the central $\rho_B$ is 
always less than the corresponding $\rho$. For example, the white dwarf with $B_{\rm cent}=4.4\times10^{15}$ G 
having mass $2.39M_\odot$ has a central $\rho_B=8.6\times10^8$ gm/cc while $\rho$ 
at the center is $4.1\times10^9$ gm/cc. Hence, depending on the magnetic field profile inside the white dwarf, the 
field can decay such that $\rho_B$ always remains sub-dominant compared to $\rho$.
For the one-Landau level system with maximum Fermi energy $E_{Fmax}=200m_ec^2$ considered 
in \cite{prl13}, the {\it central} magnetic field is $8.8\times 10^{17}$ G and the 
corresponding {\it central} magnetic density is $3.4\times10^{13}$ gm/cc. Note that 
the corresponding central $\rho$ due to the degenerate matter is $1.1\times10^{13}$ gm/cc, 
which is only a factor of $3$ smaller than $\rho_B$ at the center. 
However, both $\rho$ and $\rho_B$, as mentioned above, 
fall off away from the center of the white dwarf (as was also argued in \cite{prd12}); in fact, 
away from the center again $\rho_B$ will be 
sub-dominant compared to $\rho$. Although there will be no magnetic pressure gradient in the 
central region (due to constancy of $B$), 
as $B$ starts falling away from the center, a non-zero magnetic pressure gradient 
appears.
Note that, while the effect of $\rho_B$ increases the mass of the white dwarf at a given
radius, simultaneously it also increases gravitational power rendering the decrease of radius and 
the effect of a corresponding pressure gradient further renders the increase of radius. It is this
combined effect of $B$ in the various terms of the equations, which determines the final mass and 
radius of the white dwarf. 
We will address the stability issue arising due to the various competing effects 
in \S\ref{stability}.

However, some authors \cite{rufini} have completely missed the above point and 
have incorrectly argued that for the white dwarf having $B_{\rm cent} = 8.8\times 10^{17}$ G and $R=70$ km, 
the contribution of $B$ to the gravitational mass will be $\sim 24M_\odot$, assuming 
that the field is constant throughout the white dwarf.
If for the sake of argument we indeed consider a constant 
$B$ throughout the above white dwarf, then in addition to the contribution 
from $\rho_B$ to total density in the mass equation, as
the authors \cite{rufini} argue, there will be 
a non-zero contribution of $\rho_B$ along with $\rho$ in the force balance equation as well,
with a zero magnetic pressure gradient. If one 
simply solves these two equations simultaneously again, then one would find that both the mass and radius of the 
white dwarf decrease drastically, due to the increased gravitational force. The mass of such a 
white dwarf becomes $0.099 M_\odot$ corresponding to a radius $10.5$ km, 
which is consistent with the mass obtained from the product of $\rho_B$ and the 
volume of this white dwarf. Therefore, the previous authors \cite{rufini} should have consistently computed
the mass of the white dwarf (even if they wanted to assume the existence of a constant $B$ throughout
the star), which is a matter of a very simple numerical computation, before making such a
drastic statement based on {\it philosophy}. Figure 1 shows the 
variation of mass and pressure with radius inside the white dwarf for two cases --- 
one with $\rho_B$ and another without. 
Thus, if one simply multiplies $3.4\times10^{13}$ gm/cc by the {\it total} volume 
of the white dwarf with radius $70$ km (which was obtained without taking
$\rho_B$ into account), one arrives at an incorrect mass of $\sim 24M_\odot$. 
The more realistic case is however, where $B$ decreases with radius 
inside the white dwarf, as mentioned above. In that case we note again that, one cannot simply 
multiply $\rho_B=3.4\times10^{13}$ gm/cc by the total volume of the white dwarf, as not only does 
this $\rho_B$ pertain to only the (smaller) central region of the white dwarf, but also one must self-consistently 
account for the effects of the magnetic pressure gradient and $\rho_B$ on to the gravitational force 
and hence on the mass and radius of the white dwarf, if one
intends to include the effect of $\rho_B$ at all. All the above arguments are true for a lower $B_{\rm cent}$ as well.
Hence, the varying $B$ might lead to an increase of mass and radius again.
It is, however, still possible that even then the total mass of the white dwarf will be restricted to below 
$4.67M_\odot$ --- a mass limit proposed as a consequence of 
varying $B$ in our recent work \cite{grf13}

\begin{figure*}
\begin{center}
\includegraphics[angle=0,width=15cm]{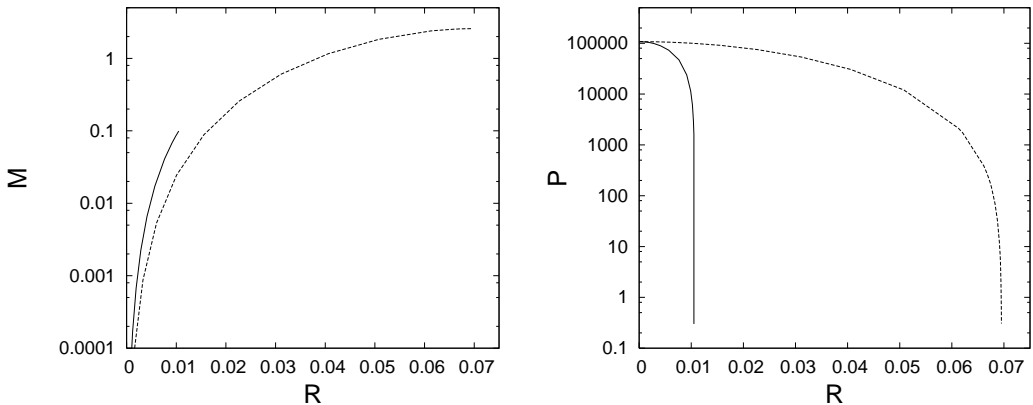}
\caption{The left panel shows mass $M$ as a function of radius $R$ and the right panel shows pressure $P$ 
as a function of radius $R$, within a white dwarf with central magnetic field 
$8.8\times10^{17}$ G. In both the panels, the solid line represents the case with a constant magnetic density of 
$3.4\times 10^{13}$ gm/cc throughout the white dwarf and the dashed line represents the case without considering magnetic density.
$R$ is in units of $10^8$ cm, $M$ in units of $M_{\odot}$ and $P$ in units of $2.668\times 10^{27}$ erg/cc. }
\end{center}
\end{figure*}

An important point to be kept in mind is that, all the 
above discussion is still within the Newtonian framework. If one truly needs 
to understand the effect of $\rho_B$, then one should carry out 
general relativistic calculations before coming to a conclusion.
In this connection, let us admit that we in principle also should have conducted 
a general relativistic treatment. However, we wanted to keep the work 
in the spirit of Chandrasekhar's calculation and accessible to the general readers.
Note that for $B = 8.8\times 10^{15}$ G, the mass of the white dwarf ($\sim 2.4M_\odot$), although 
has not reached the proposed limiting mass ($\sim 2.58M_\odot$) yet, is just slightly away from it, 
while the white dwarf is still in the Newtonian regime. Therefore, our computation of
limiting mass, in the Newtonian framework itself, should not be far from exact.

\section{Stable equilibrium in the presence of strong magnetic field}
\label{stability}

One might wonder (as raised in \cite{rufini,rajaram}) whether magnetic energy could 
dominate over internal energy without magnetic forces significantly
altering the equilibrium. Since the central region of the white dwarf in 
our works \cite{kundu,prd12,ijmpd,prl13,apjl13,grf13} is modeled to have a constant $B$, the magnetic forces 
vanish therein. Nevertheless, the field must drop off by the time the
surface is reached and hence the field gradients (and
therefore magnetic forces) could exist primarily in the lower density
outer regions, which raises the question of stability. 
We address this with the help of two arguments.

First, let us consider a simplistic magnetic field profile for the white dwarf 
such that the field is constant in a central region and 
then varies away from the center such that the components of the field 
in spherical polar co-ordinates are $(B_r,0,0)$, where 
$B_r \propto \cos \theta/r^2$. 
Note that $\nabla \cdot \vec{B} = 0$ for such a field configuration. 
Such a profile physically corresponds to the field lines coming out of one hemisphere and 
going through the other hemisphere of the star.
Now, in the presence of a 
varying $B$ the white dwarf must obey the condition of magnetostatic equilibrium:
\begin{equation}
\frac{1}{\rho+\rho_B}\frac{d}{dr}\left(P+\frac{B^2}{8\pi}\right)=-\frac{GM}{r^2}+\left[\frac{(\vec{B}\cdot\nabla)\vec{B}}{4\pi(\rho+\rho_B)}\right]_r,
\label{magstat}
\end{equation}
where, $r$ is the radial distance from the center of the white dwarf and $P$ is the pressure of the electron degenerate matter. 
Interestingly, for the above mentioned field configuration, 
the magnetic pressure gradient term on the left hand side and 
the magnetic tension term on the right hand side of equation (\ref{magstat}) cancel each other,
and one is simply left with the equation for hydrostatic equilibrium.
However, in reality they never exactly cancel each other, giving rise to a net effect of magnetic 
pressure gradient which will render a larger radius with respect to that in its absence. 
Now as to the effect of $\rho_B$, 
although it is constant in a smaller central region, away from the center it becomes 
proportional to $\cos^2 \theta/r^4$, which falls off so fast
that in the outer region it has practically no contribution. The constant $B$ and then $\rho_B$ in the center, 
however, will shrink the mass and 
radius in the central region which will affect the overall mass and radius of the white dwarf.
Hence, the combined effect is likely to produce the result obtained in our works \cite{prd12,prl13}. 
Note that, as shown in \S\ref{equi}, a non-zero gradient of magnetic pressure can lead to a factor of $4.4$ mismatch 
between $B_{\rm avg}$ and $B_{\rm equi}$ of the white dwarf, which is nothing astrophysically, when there are so
many uncertainties involved. Moreover, recalling that $B_{\rm equi}$ is just an 
``of the order" estimate, previous authors \cite{rufini,rajaram} 
should not come to a rigid conclusion against the stability of our model based simply on that.
In all these cases the white dwarf will no longer remain spherical but may attain stability 
by assuming an oblate spheroidal shape. However, the oblate spheroidal white dwarf
will become super-Chandrasekhar at a much smaller $B$ compared to its spherical 
counterpart (see \cite{prd12}), which might automatically satisfy the stability criteria. Hence,
in a more self-consistent computation, one may not need to invoke such a higher field in order to
approach the limiting mass.

Next, we consider a scenario where $B$ is fluctuating with radius within 
the white dwarf. We quote from the work of Broderick et al. \cite{broderik}, which states that 
``It is even possible to imagine a disordered field where $<B^2>$ is significantly larger 
than $<\vec{B}>^2$." 
In a similar context, Suh and Matthews \cite{suh} have shown that magnetization of 
magnetar-matter due to strong magnetic field leads to the formation of magnetic domains. 
Cheoun et al. \cite{jcap} have invoked such randomly oriented magnetic domains to 
justify the spherically symmetric magnetic field configuration of the strongly magnetized neutron stars 
in their work. They \cite{jcap} further say that within the region of domain formation, the random currents generated 
due to the randomly oriented magnetic fields would cancel each other leading to a net zero current. 
This implies that the domain forming region consists of force free fields. Thus, taking all these 
into consideration, it is quite plausible to think of a {\it tangled} field 
configuration, especially in the central region of the white dwarf.
Note that, the length scale over which 
the field can fluctuate ($l_f$) must be much larger than the quantum mechanical length scale ($l_{qm}$) characteristic 
to Landau quantization, given by the magnetic length $l_{qm} = (\hbar c/eB)^{1/2}$. For the case with a strong  
$B_{\rm cent}=8.8\times10^{17}$ G, it turns out that $l_{qm} = 2.7\times 10^{-13}$ cm, 
which is likely to be much smaller than any spatial variation in the stellar magnetic field. Thus, inside the 
corresponding white dwarf with radius $\sim 10^7$ cm, it is 
quite reasonable to believe that $l_{qm} \ll l_f \ll l_{mhd}$, where $l_{mhd}$ is the length scale over which the 
fluid approximation is valid such that the magnetostatic equilibrium equation can be solved. 
In such a situation, the electron gas is Landau quantized due to the strong local central field, however 
the average field of a single fluid element is much smaller due to the fluctuations of the field within it. 
Since the white dwarf constitutes of numerous such fluid elements, it has a net low (global) average field, 
which implies a much lower $\rho_B$ and also lower field pressure. Hence, the gradient 
of magnetic pressure in the magnetostatic balance equation will also be significantly lower. 
Towards the surface of the white dwarf the magnetic field strength starts decreasing and, hence, although $l_{qm}$ 
increases, still it remains much smaller compared to $l_f$ (and the field gradient 
itself is lower therein), thus attaining a stable configuration.
Therefore, depending on 
how fast $B$ is fluctuating from a large central amplitude to a smaller surface amplitude, the 
(global) average field of the white dwarf can become much lower than the corresponding $B_{\rm equi}$, 
thus maintaining the condition for dynamical stability.

In a related context, some authors \cite{chamel} have suggested that 
the onset of neutronization and pycnonuclear reactions may limit the stability of 
the concerned super-Chandrasekhar white dwarfs, due to a softening of the underlying
EoS. We recall here that a similar softening is brought about when a 
constant $\rho_B$ is added to $\rho$, without changing the pressure, while solving for the structure 
of the white dwarf. This 
leads to a decrease in the mass and radius of the white dwarf, due 
to the increased inward gravitational pull, as described in \S\ref{rhoB} (also see Figure 1). 
A similar result is apparent in the presence of neutronization and pycnonuclear reactions.
However, this is equally valid for nonmagnetized Chandrasekhar's white dwarfs whose
central $\rho$ is $\sim 10^9$ gm/cc as well.
The authors \cite{chamel} further show that for 
the white dwarfs to be stable against neutronization, $B$ at the 
center should be less than few times $10^{16}$ G for typical matter compositions. 
Note that, as mentioned before at the end of \S\ref{rhoB}, for $B_{\rm cent} \lesssim 10^{16}$ G the mass-radius 
relation already starts approaching the limiting mass, yielding super-Chandrasekhar white dwarfs with 
mass $\gtrsim 2.44M_{\odot}$, in the Newtonian regime itself. Clearly these white 
dwarfs are well within the threshold for the onset of neutronization. 
As to the rates of pycnonuclear reactions, they are very uncertain, as the authors \cite{chamel} themselves admit, 
and hence it is probably premature to set a bound on the maximum allowed $B$ based on that. 
More importantly, one cannot rule out the possibility that a limiting mass can be obtained at a 
field well within the limits set by various stability criteria, once a more accurate, self-consistent
calculation is performed.

\section{Revisiting limiting mass with varying magnetic field}

In this section, we would like to re-explore the limiting mass, which is at the 
extreme relativistic limit, by considering the effect of both the magnetic pressure gradient and 
$\rho_B$.

By referring to the supplementary material 
of \cite{prl13}, we show below that at the extreme relativistic limit (i.e. Fermi energy $E_F \gg m_e c^2$ and 
$B_{\rm cent}\gtrsim 10^{17}$ G)
the plasma-$\beta$ ($ = P/P_B$), with magnetic pressure $P_B=B^2/8\pi$, 
becomes independent of $B$. 
Let us concentrate on this regime now. 
However, as mentioned before, although a large $B$ brings in anisotropy
in the system, which may lead the star to become spheroidal, making a general relativistic treatment
indispensable, for the present purpose we plan to stick to the Newtonian framework assuming the star
to be spherical. Therefore, we consider the overall isotropic magnetic pressure to be $P_B=B^2/24\pi$. 
In this case, the matter pressure for a one-Landau level system reduces to 
\begin{equation}
P = \frac{B_D m_e c^2}{4 \pi^2 \lambda_e^3} \left(\frac{E_F}{m_e c^2}\right)^2 ,
\label{pres1lev}
\end{equation}  
where for all the symbols please 
refer to \cite{prl13} and its supplementary material.
Since, in the extreme relativistic regime, for a one-Landau level system
\begin{equation}
(E_F/m_e c^2)^2 \approx 2 B_D , 
\label{ef_bd}
\end{equation}
equation (\ref{pres1lev}) becomes
\begin{equation}
P = \frac{B^2 m_e c^2}{2 \pi^2 \lambda_e^3 B_c^2} 
\end{equation}
and thus we obtain
\begin{equation}
\beta = \frac{12 m_e c^2}{\pi \lambda_e^3 B_c^2} \approx 0.028 .
\end{equation}
Hence, $P_B \approx 35.7 P$, which is independent of $B$.
Now we recall the magnetostatic balance equation (without the magnetic tension term) 
including $\rho_B$ and the equation for the estimate of mass:
\begin{equation}
\frac{dP}{dr} + \frac{dP_B}{dr} = -\frac{G M(r)}{r^2} (\rho + \rho_B),\,\,\,\,\frac{dM}{dr}=4\pi r^2(\rho+\rho_B) .
\end{equation}
Hence, for the extreme relativistic case the above set of equations reduces to 
\begin{equation}
36.7 \frac{dP}{dr} = -(1+X)\rho \frac{GM(r)}{r^2} ,\,\,\,\,\frac{dM}{dr}=4\pi r^2\rho(1+X),
\label{magneto}
\end{equation}
where $X = \rho_B/\rho$, 
\begin{equation}
\rho = \frac{\mu_e m_H B_D E_F}{2 \pi^2 m_e c^2 \lambda_e^3} .
\end{equation}
Using equation (\ref{ef_bd}) the above equation reduces to 
\begin{equation}
\rho = \frac{\sqrt{2} \mu_e m_H B_D^{3/2}}{2 \pi^2 \lambda_e^3} .
\end{equation}

Now, it is known that a magnetized compact object will become a black hole at $B\sim 10^{19}$ G,
depending on its mass and spin (see \cite{cardal}). 
For this value of $B$, $X=9.9$. Therefore, by solving the set of equations (\ref{magneto}) we obtain $M\approx4.75M_\odot$
--- very similar to that predicted by our recent work \cite{grf13}, when $B$ is varying inside the white dwarf.
Nevertheless, we would like to emphasize again that this is a rough estimate (motivated by Chandrasekhar's
idea) to get the basic picture.
In reality, this has to be done in the framework of general relativity. In addition, the above $\beta$
will remain constant only in the central region of the white dwarf.
Away from the center it should change according to the 
variation of $P$ and $P_B$. For example, near the surface of the white dwarf, $P_B$ becomes 
much subdominant compared to $P$ and hence $dP/dr>>dP_B/dr$ 
and $X$ to decrease significantly.


\section{Stability against magnetic buoyancy}

\begin{figure*}
\begin{center}
\includegraphics[angle=0,width=17cm]{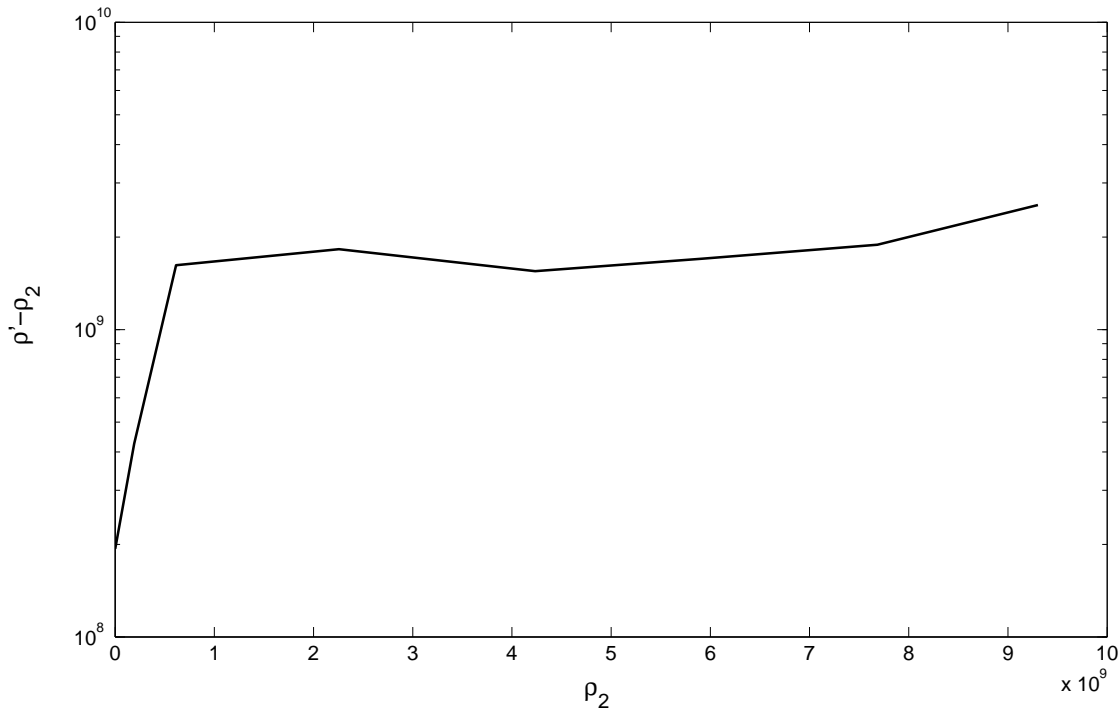}
\caption{Variation of the difference in densities inside and outside a displaced blob of gas 
as a function of its surrounding density. Density is in units of gm/cc.}
\label{buyo}
\end{center}
\end{figure*}

A question that may arise regarding the issue of buoyancy related instabilities within the strongly
magnetized super-Chandrasekhar white dwarfs. 
If a blob of gas is displaced adiabatically from an inner radius to an
outer radius in the central region of the white dwarf,
then the condition for stability is given by
\begin{equation}
\rho' =\left(\frac{K_1}{K_2}\rho_{1}^{\Gamma_1} \right)^{1/\Gamma_2} > \rho_2,
\end{equation}
where $\rho'$ is the matter density inside the blob when it is at the outer radius,
$\rho_1$, $\Gamma_1$ and $K_1$ are the matter density, adiabatic index and a dimensional constant
related to the adiabatic EoS respectively of the surrounding medium at the inner 
radius and $\rho_2$, $\Gamma_2$ and $K_2$ are the same quantities at the outer radius.
Note that, since $B$ is constant in the central region
of the white dwarf, the magnetic pressure terms do not appear in the above
relation. The above criterion means that the blob at the outer radius must be heavier than
the surrounding (i.e., $\rho'-\rho_2 > 0$) so that it is pulled inwards by gravity to its
original position, making the system stable.
If the blob is lighter than its surrounding (i.e., $\rho'-\rho_2 < 0$) then the buoyant
force displaces the blob further outwards, making the system unstable.

We apply this argument to a white dwarf having
central density $\sim 1.4\times 10^9$ gm/cc; the EoS of the
underlying degenerate gas being described by the parameters listed in Table II of
\cite{prd12}. In Figure 2 we show the variation of $\rho'-\rho_2$ with $\rho_2$ inside
this white dwarf described by various zones with piecewise constant $\Gamma$. We clearly
see that $\rho'-\rho_2$ is everywhere positive, establishing that the
white dwarf is stable against buoyancy related instabilities.
Away from the central region, $B$ decreases
towards the surface (as discussed previously),
so the magnetic pressure at the outer radius will be less than that at the inner radius. Taking
this difference into account will only increase the density of the blob, making it even more
heavier than the surrounding, further strengthening the argument for a stable system.

\section{Difficulty behind direct observation of strongly magnetized super-Chandrasekhar white dwarfs}

Recent observations of several peculiar over-luminous type Ia supernovae - SN 2006gz, SN 2007if, 
SN 2009dc, SN 2003fg - seem to suggest super-Chandrasekhar-mass white dwarfs with masses in the range $2.1-2.8M_{\odot}$ 
as their most likely progenitors \cite{howell,scalzo}. 
In order to the explain these observations, some authors \cite{hachisu} have performed 
simulations of accreting binary white dwarfs in a single degenerate scenario, which include 
the effect of differential rotation and other phenomena.
Models have also been proposed which invoke the merger of two white dwarfs, known as the double degenerate
scenario \cite{scalzo}.
On a fundamentally different ground, the strongly magnetized super-Chandrasekhar white dwarfs proposed by Das \& 
Mukhopadhyay \cite{prd12,prl13} and Das, Mukhopadhyay \& Rao \cite{apjl13} can explain these supernova 
explosions, which is the indirect evidence for the existence of super-Chandrasekhar white dwarfs.
However, there has been no direct detection of 
such white dwarfs so far. We try to explore the reason behind this in this section.

Ordinary sub-Chandrasekhar white dwarfs are located 
in the lower-left portion of the Hertzsprung-Russell (H-R) diagram (plot showing the relationship 
between the luminosities of stars and their effective temperatures), indicating that they are low luminosity hot 
objects. If the surface of the white dwarf is considered to radiate like a black body, then its luminosity is given by 
$L = 4\pi R^2 \sigma_B T_{eff}^4$, where $\sigma_B$ is the Stefan-Boltzmann constant and $T_{eff}$ the 
effective surface temperature of the white dwarf. Since $L\propto R^2$, ordinary white dwarfs, 
having radius $\sim 10^4$ km, are bright enough to feature in the H-R diagram. Neutron stars on the 
other hand, having radius $10$ km, are too small (and hence too dim) and would lie far below the white dwarfs in the 
H-R diagram, and hence do not appear in it altogether. Note that, 
neutron stars are mostly detected as pulsars, due to the emissions of regular pulses.
The strongly magnetized super-Chandrasekhar white dwarfs in our works, having radius $\lesssim 100$ km, 
are more closer to neutron stars than conventional white dwarfs. Thus, such (isolated) white dwarfs 
would also {\it not} appear in the H-R diagram, making them virtually impossible to be detected directly 
based on current observational techniques.

\section{Separate flux conservation in the central and surface regions}

Das, Mukhopadhyay \& Rao \cite{apjl13} have discussed a possible evolutionary scenario from 
weakly magnetized, sub-Chandrasekhar, accreting white dwarfs to strongly magnetized, 
super-Chandrasekhar white dwarfs, which eventually explode to give rise to the peculiar 
type Ia supernovae. Irrespective of how the interior magnetic field of the white dwarf 
is modeled, the only constraint on the above evolution is the magnetic flux conservation between 
the initial and final white dwarfs, which controls/relates the magnetic field profiles 
of the white dwarfs, if one of them (say the initial one) is known. Additionally, 
the authors have imposed that the magnetic flux is conserved separately in the central 
and surface regions of the white dwarfs. This is justified as follows.

In \S2 of \cite{apjl13}, the authors have defined a radius $R_{int}$ which demarcates 
a zone up to which the central magnetic field of the white dwarf remains (almost) 
constant, which is larger than that at the surface. 
This is quite a plausible assumption, since the original star, which 
collapses to form a white dwarf, might have a very large interior magnetic field 
compared to that observed on its surface. For example, Parker \cite{parker} 
estimated that a magnetic field strength as high as $\sim 10^7$ G can not be ruled out for the 
Sun's interior. Gough \& McIntyre \cite{gough} also discussed the inevitability of 
a strong magnetic field in the Sun's interior and its influence to the extent 
of a transitional layer, called the tachocline. Thus, it is quite likely that the values 
of the central and surface magnetic fluxes (in the initial star and hence in the 
white dwarf) and the underlying electric currents might be significantly different. 
Therefore, the size of the central (and
then the surface) region of a white dwarf is controlled by the size of the central
region of the original star. Since it is likely that there is a gradient of magnetic 
field from the center to the surface of the star, the radial dependence 
of $B$ could be of the form $B\propto 1/r^\zeta$, where $\zeta$ is positive. 
However, the above field profile will blow up at $r=0$ which is unfeasible and hence 
$B$ is expected to be very large but to remain constant in the central region.
This restricts the electrons to move around in the central region of the 
final highly magnetized white dwarfs, as only
the lower Landau level(s) is/are available for them, justifying practically no 
current therein. Nevertheless, as $B$ decreases
towards the surface, higher Landau levels become available and hence the electrons
can produce current, maintaining a gradient of the magnetic field up to the surface.

All the above arguments justify the assumption of the separate conservation of 
the surface and central magnetic fluxes, which also suggest the presence of 
a core-envelope boundary.

\section{Correlation between the observed peculiar type Ia supernovae rate and 
the occurrence of highly magnetized super-Chandrasekhar white dwarfs}

Identification of the progenitors of type Ia supernovae is still an ongoing research area. 
In \cite{apjl13} we have proposed
that if the progenitors are single degenerate white dwarfs accreting 
matter from a normal companion, then the peculiar supernovae originate from super-Chandrasekhar white dwarfs 
with extremely high magnetic fields. An underlying assumption for such a scenario is that 
the matter-supplying companion star should be massive enough ($\gtrsim 3M_\odot$) for the white dwarf 
to accrete enough mass and become super-Chandrasekhar (as also assumed, for example, in \cite{hachisu}).

The observed rate of over-luminous, peculiar type Ia supernovae is about 
1-2\% of the normal type Ia \cite{scalzo12}. 
This is consistent with the fact that about 25\% 
of all cataclysmic variables (CVs) are magnetic CVs and we expect a small fraction 
(say 10\%) of them to be super-magnetic. 
These high magnetic CVs (2.5\% of all CVs based on the above assumption) would 
behave like normal magnetic CVs during most of their life time, till their mass 
remains sub-Chandrasekhar. However, they would be seen as peculiar objects only 
during a small fraction of their life time, when $B_{\rm surf}$ becomes 
very high ($>10^{11}$ G) and the mass is super-Chandrasekhar ($> 1.44 M_\odot$), 
which should happen because of accretion. We have discussed this issue with constant and varying 
accretion rates in \S3 of \cite{apjl13}.


For a constant accretion rate of $2.5\times 10^{-9} M_\odot {\rm yr^{-1}}$, 
which leads to a white dwarf having final 
mass $2.33M_\odot$ (shown by the dashed line in Figure 3(a) of \cite{apjl13}), the white dwarf 
spends about 0.4 fraction of its lifetime with super-Chandrasekhar mass. For varying accretion rate, 
modeled by equation 4 of \cite{apjl13} with $\alpha=2$ (solid line in Figure 3(b) of \cite{apjl13}), the 
white dwarf spends about 0.05 fraction of its lifetime with super-Chandrasekhar mass.
Now, typically observed stably accreting intermediate polars have accretion 
rates $\sim 10^{-9} M_\odot {\rm yr^{-1}}$ \cite{warner}. However, an accretion rate 
$\gtrsim 10^{-7} M_\odot {\rm yr^{-1}}$ is required for the 
stable burning of the accreted hydrogen and helium in (single degenerate) 
type Ia supernovae progenitors \cite{cumming}, which is indeed attained by the above
model with varying accretion rate, starting with an accretion rate similar to that of
intermediate polars. Thus, the varying accretion rate is a more likely scenario, 
which gives us the estimate that about 0.12 \% of CVs would be peculiar high magnetic CVs. 

Since these white dwarfs are 
accreting at a high rate, they will be visible as bright X-ray objects, rather than white dwarfs in 
optical data bases like SDSS.
Interestingly, the recent MAXI observation of an unusually luminous 
soft X-ray transient, MAXI J0158−744, which is a binary system consisting 
of a white dwarf and a
Be star \cite{morii}, seems to support our claim. In order to explain the 
puzzling super-Eddington X-ray outburst from this source, 
Morii et al. \cite{morii} conclude from their analysis 
that the underlying white dwarf must have a mass near the Chandrasekhar limit, or exceed it considerably 
and must also be highly magnetized, as proposed by us \cite{kundu,prd12,ijmpd,prl13,apjl13,grf13}.

\section{Summary}

We have reestablished the subtle physics issues related to highly magnetized 
super-Chandrasekhar white dwarfs. We have shown that such white dwarfs are 
indeed possible, provided certain obvious physics constraints 
are satisfied, explained in this paper. This further strengthens the argument 
for the existence of such super-Chandrasekhar white dwarfs.

Although it is based on some simplified assumptions in the first place, our work 
attempted to open a new window in modern astrophysics.
We must admit that in reality, the effect of general relativity,
magnetic field profile etc. have to be included in the model self-consistently.
However, the series of works \cite{kundu,prd12,ijmpd,prl13,apjl13,grf13} we are revisiting is just the initiation. 
All of them were done in the spirit of Chandrasekhar's work. 
More self-consistent computations may turn out to give deviated results.
It also maybe possible that the limiting mass, which we have proposed,
can be obtained at a lower field, once all the effects are
taken into account properly (e.g. in general relativity the
limiting mass does not correspond to zero radius). Nevertheless,
one of the main claims that significantly super-Chandrasekhar
white dwarfs can exist which can explain peculiar, over-luminous
type Ia supernovae, corresponding to a new limiting mass
(irrespective of its exact value), remains intact. Now it is
the responsibility of the community to come forward to make
the idea more concrete, in place of attempting to nip it
at the bud.

\section*{Acknowledgments}

The authors would like to thank Jeremiah Ostriker of Princeton University for encouraging 
comments and appreciation. The authors also acknowledge A. R. Rao of 
TIFR, Diptiman Sen, Tarun Deep Saini and Prateek Sharma of IISc for discussion and encouragement. 
Thanks are further due to Rajaram Nityananda of IISER-Pune for his open criticisms.
This work is partly supported by ISRO project Grant No. ISRO/RES/2/367/10-11.
U.D. thanks CSIR, India for financial support.

\end{document}